\documentclass[11pt]{article}
\usepackage[hyperref]{ccl2023-en}
\usepackage{url}
\usepackage{latexsym}
\usepackage{fancyhdr}

\usepackage{graphicx}
\usepackage{amsmath}
\usepackage{booktabs}
\usepackage{multirow}
\usepackage{tabularx}
\newcolumntype{R}{>{\raggedleft\arraybackslash}X}
\usepackage{subcaption}
\usepackage{enumitem}

\usepackage{cleveref}

\title{Distinguishing Neural Speech Synthesis Models Through Fingerprints in Speech Waveforms}

\author{Chu Yuan Zhang$^{1,2}$, Jiangyan Yi$^1$, Jianhua Tao$^{3,4}$, Chenglong Wang$^1$, Xinrui Yan$^{1,2}$\\[1ex]
$^1$ Institute of Automation, Chinese Academy of Sciences, Beijing, China\\
$^2$ University of Chinese Academy of Sciences, Beijing, China\\
$^3$ Department of Automation, Tsinghua University, Beijing, China\\
$^4$ Beijing National Research Center for Information Science and Technology,\\Tsinghua University, Beijing, China
}

\date{2024}

\begin{document}
\maketitle
\begin{abstract}
  Recent advancements in neural speech synthesis technologies have brought about widespread applications but have also raised concerns about potential misuse and abuse. Addressing these challenges is crucial, particularly in the realms of forensics and intellectual property protection. While previous research on source attribution of synthesized speech has its limitations, our study aims to fill these gaps by investigating the identification of sources in synthesized speech. We focus on analyzing speech synthesis model fingerprints in generated speech waveforms, emphasizing the roles of the acoustic model and vocoder. Our research, based on the multi-speaker LibriTTS dataset, reveals two key insights: (1) both vocoders and acoustic models leave distinct, model-specific fingerprints on generated waveforms, and (2) vocoder fingerprints, being more dominant, may obscure those from the acoustic model. These findings underscore the presence of model-specific fingerprints in both components, suggesting their potential significance in source identification applications.
\end{abstract}

\section{Introduction}
\label{intro}

%
%
\cclfootnote{
    %
    %
    \hspace{-0.65cm}  
    \textcopyright 2024 China National Conference on Computational Linguistics

    \noindent Published under Creative Commons Attribution 4.0 International License
}

Recently, neural speech synthesis (also known as text-to-speech, or TTS) systems have made substantial advancements in generating highly realistic speech waveforms. These neural TTS systems can mainly be divided into two categories. The first category, pipeline systems, generally comprise an acoustic model such as Tacotron~2~\cite{shen2018tacotron2}, FastSpeech~2~\cite{ren2021fastspeech} and GradTTS~\cite{popov2021gradtts}, alongside neural vocoders such as Parallel WaveGAN~\cite{yamamoto2020parallel}, HiFiGAN~\cite{kong2020hifigan}, Style MelGAN~\cite{mustafa2021stylemelgan} and Multiband MelGAN~\cite{yang2020multiband}. The second category, end-to-end systems, also have been gaining traction in recent years. These systems, such as FastSpeech~2s~\cite{ren2021fastspeech} and VITS~\cite{kim2021vits}, are capable of generating speech waveforms directly from text.

While these advancements have found widespread application, they raise concerns about misuse in fraudulent activities, threatening security and privacy. Consequently, researchers have aimed to develop robust methods for differentiating genuine human speech from synthetic counterparts~\cite{yi2023audio}. Initiatives like the ASVspoof challenges~\cite{wu2015asvspoof,kinnunen2017asvspoof,wang2020asvspoof,yamagishi2021asvspoof} and the ADD challenges~\cite{yi2022add} showcase advancements in audio anti-spoofing, including the use of front-end spectral features like linear frequency cepstral coefficients (LFCCs)~\cite{todisco2018integrated} as well as backend classification models like Res2Net~\cite{gao2021res2net} in anti-spoofing.

\begin{figure}
	\centering
	\includegraphics[width=0.8\linewidth]{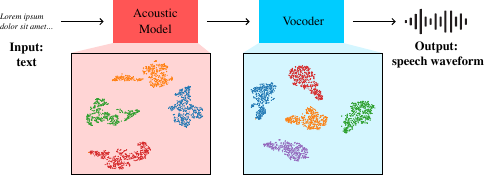}
	\caption{TTS pipeline and t-SNE projection of acoustic model and vocoder fingerprints.}
	\label{fig:main}
\end{figure}

\begin{figure}[t]
	\centering
	\includegraphics[width=0.9\linewidth]{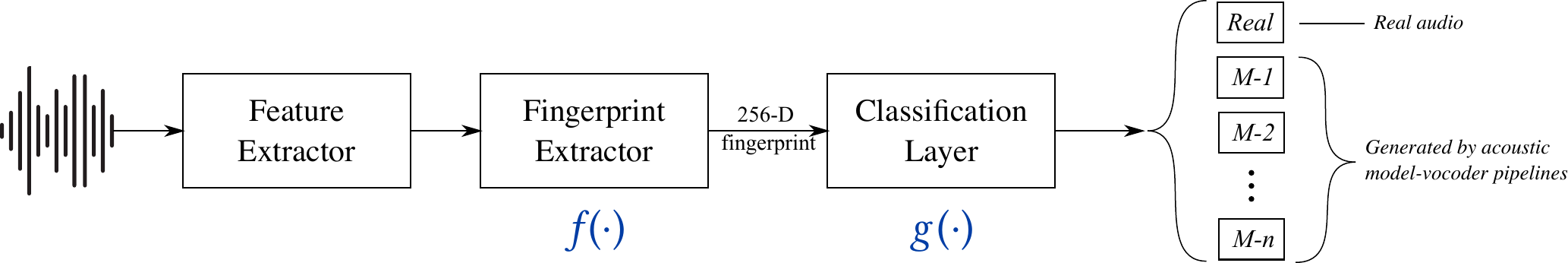}
	\caption{Fingerprint extraction pipeline.}
	\label{fig:extraction}
\end{figure}

There is a also growing interest in identifying the source of synthesized speech, especially in forensics and intellectual property protection. The ADD 2023 challenge~\cite{yi2023add} includes a deepfake algorithm recognition track, addressing the recognition of synthesis tools, where various methods have been proposed~\cite{lu2023detecting,tian2023deepfake}.
It is however worth noting that deepfake source attribution is a research topic that has not been extensively explored. Pons \emph{et al.}~\cite{pons2021upsampling} focused on upsampling artifacts but limited their investigation to visual representation without exploring exploitability for attribution.
Similarly, Yan \emph{et al.}~\cite{yan2022initial} studied vocoder fingerprints but overlooked finer parameter differences within vocoders of the same architecture. 
This underscores the need for more comprehensive investigations covering various factors to enhance our understanding of deepfake source attribution. Notably, the traditional approach to TTS usually involves a pipeline system, where an acoustic model generates mel spectrograms from text, which are then converted into speech waveforms by a vocoder; mainstream approaches to source attribution have largely overlooked the potential of these components to leave fingerprints on the synthesized speech by performing attribution on a pipeline-level basis.

In the interest of filling this gap and fostering a more robust and nuanced comprehension of deepfake source attribution, we present our study. We aim to extract fingerprints from synthesized speech to identify their sources, focusing on acoustic model and vocoder fingerprints in pipeline TTS systems. In particular, we put forth and seek to explain the following questions: (1) Do acoustic models and vocoders leave model-speciﬁc ﬁngerprints on the speech waveforms they generate? (2) When taking into consideration both the acoustic model and the vocoder, does one overshadow the other in terms of fingerprints? 

To answer these questions, we train various acoustic models and vocoders on the LibriTTS~\cite{zen2019libritts} dataset to form TTS pipelines. 
Through waveform fingerprint extraction from these pipelines, we assess our hypothesis that both the acoustic model and vocoder leave model-specific fingerprints on the generated waveforms. Our experiments confirm these fingerprints' existence, highlighting the vocoder's more prominent role, potentially masking acoustic model fingerprints. These findings suggest the potential utility of model-specific fingerprints in source identification applications.

In this paper, we present our methodology in Section~2, followed by our experimental setup and results in Section~3, and then conclude with a discussion of our findings in Section~4.

\section{Methodology}

In this section we present the methodology employed for the verification of fingerprints within synthesized speech generated by TTS pipelines. Abstractly, a speech synthesis pipeline composed of a given acoustic model $a(\cdot)$ and a given vocoder $v(\cdot)$ can be viewed as a function that maps a text sequence \(x\) to a speech waveform \(y\), i.e., \(y = v(a(x))\). Suppose there theoretically exist a ``perfect'' acoustic model \(a^*\) and a ``perfect'' vocoder \(v^*\) that generate ``perfect'' speech waveforms that are mathematically indistinguishable from genuine human speech. As current TTS pipelines are imperfect, with imperfections arising from both the theoretical foundation and the practical implementation, we can therefore expect that the speech waveforms generated by these pipelines will deviate from perfection; i.e.,
\begin{equation}\label{eq:method:premise}
	\begin{aligned}
		a(x) & = a^*(x) + \epsilon_a(x) \\
		v(a(x)) & = v^*(a(x)) + \epsilon_v(a(x))
	\end{aligned}
\end{equation}
where $\epsilon_a$ and $\epsilon_v$ are residuals that contain the fingerprints of the acoustic model and the vocoder, respectively.

Inspired by the assumption in \newcite{marra2019gans}, we assume that the fingerprints of the acoustic model and the vocoder are independent of each other, and that each of the residuals $\epsilon_a$ and $\epsilon_v$ contains the fingerprint plus a Gaussian noise, such that, with a sufficiently large dataset, the Gaussian noise can be ignored. We therefore seek to verify the existence of such fingerprints through an extraction network \(f(\cdot)\) that extracts the fingerprints from the speech waveform \(y\), i.e., either
\begin{equation}
  \begin{aligned}
    f_a(y) &= \hat{\epsilon}_a(x)\\
    f_v(y) &= \hat{\epsilon}_v(a(x))
  \end{aligned}
\end{equation}
as the case may be. We then train a classifier \(g(\cdot)\) to classify the fingerprints extracted by \(f(\cdot)\) into the corresponding acoustic model or vocoder, i.e., 
\begin{equation}
	\begin{aligned}
    g_a(f(y)) &= a \\
    g_v(f(y)) &= v
  \end{aligned}
\end{equation}
The entire extraction and classification pipeline is shown in \ref{fig:extraction}. We then investigate the performance of the classifier $g(\cdot)$ to determine whether the fingerprints of the acoustic model and the vocoder are distinguishable.

Following the methodology outlined above, we investigate the fingerprints of the acoustic model and the vocoder separately, and then investigate the relationship between the two fingerprints. We also investigate the robustness of the fingerprints against perturbations in the input speech samples.

\begin{table}
	\centering
	\small
	\caption{Vocoder model setups.}\label{tbl:vocoders}
	\begin{tabular}{lcrr}
		\toprule
		\multicolumn{1}{l}{\textbf{Architecture}} &     \multicolumn{1}{l}{\textbf{Model ID}} & \multicolumn{1}{l}{\textbf{Seed}} & \multicolumn{1}{l}{\textbf{Batch size}} \\ \midrule
		\multirow{5}{*}{Parallel WaveGAN (PWG)}&P0    &  1000 &     6 \\
		&P1    & 1001 &     6  \\
		&P2    &  1000 &     8 \\
		&P3    & 1002 & 6 \\ 
		&P4    & 1003 & 6 \\ \midrule
		\multirow{3}{*}{HiFiGAN (HFG)}&H0    &   1000 &    16 \\
		&H1    &  1001 &    16  \\
		&H2    &  1000 &     8 \\ \midrule
		\multirow{3}{*}{Multiband MelGAN (MMG)}&M0    & 1000 &    16  \\
		&M1    & 1001 &    16 \\
		&M2    &  1000 &     8 \\\midrule
		\multirow{3}{*}{StyleMelGAN (SMG)}&S0    &  1000 &    16 \\
		&S1    &  1001 &    16 \\
		&S2    &  1000 &     8  \\
\bottomrule
	\end{tabular}
\end{table}

\begin{table*}
	\centering
	\small
	\caption{Fingerprint analysis experiment setups. (Exp. = Experiment ID)\\Acoustic models used: FastSpeech~2 (F2), Grad-TTS (GD), Tacotron~2 (T2)\\Vocoders used: see \Cref{tbl:vocoders}\\In Experiments V1, V2, A1, and A2, different inputs are used with the same set of vocoders/acoustic models to generate samples for the training, validation and test sets.}\label{tbl:exp-setup}
	\begin{tabularx}{\linewidth}{cXXX}
		\toprule
		\textbf{Exp.} & \textbf{Training set} & \textbf{Validation set} & \textbf{Test set} \\ \midrule
		V1 & P0, H0, M0, S0 & (same methods as training set) & (same methods as training set) \\
		V2 & P0, P1, P3, P4 & (same methods as training set) & (same methods as training set) \\
		V3 & P0, H0, M0, S0 & P1, P2, H1, H2, M1, M2, S1, S2 & P1, P2, H1, H2, M1, M2, S1, S2 \\
		A1 & F2+H0, GD+H0, T2+H0 & (same methods as training set) & (same methods as training set) \\
		A2 & F2+H0, F2+P0, F2+S0, GD+H0, GD+P0, GD+S0, T2+H0, T2+P0, T2+S0 & (same methods as training set) & (same methods as training set) \\
		R1 & F2+H0, GD+H0, T2+H0 & F2+P1, F2+S1, GD+P1, GD+S1, T2+P1, T2+S1 & F2+P1, F2+S1, GD+P1, GD+S1, T2+P1, T2+S1 \\
		R2 & T2+P0, T2+H0, T2+M0, T2+S0 & F2+P0, F2+H0, F2+M0, F2+S0 & F2+P0, F2+H0, F2+M0, F2+S0 \\
		N1 & (R2 training set) & (R2 validation set) & (R2 test set with noise at 10 dB SNR) \\
		N2 & (R2 training set) & (R2 validation set) & (R2 test set with reverberation) \\
		N3 & (R2 training set) & (R2 validation set) & (R2 test set with speed adjustment) \\
		\bottomrule
	\end{tabularx}
\end{table*}

\subsection{Vocoder fingerprint}\label{sec:method:vocoder}

For vocoders, each model is characterized not only by the model architecture, but also by the precise weights in the neural network. This is especially true with generative adversarial networks (GANs) due to the adversarial nature of their training, meaning that the precise weights are sensitive to training setups~\cite{brock2019large}. For this reason, we investigate the impact of both model architectures and training setups in vocoder fingerprints.

To study the impact of model architectures, we train vocoders of 4 architectures: Parallel WaveGAN~\cite{yamamoto2020parallel}, HiFiGAN~\cite{kong2020hifigan}, Multiband MelGAN~\cite{yang2020multiband}, and Style MelGAN~\cite{mustafa2021stylemelgan}, and investigate the fingerprints of the generated waveforms to determine whether the architecture leaves a fingerprint. To investigate the impact of training setup, we train several vocoders of the 4 aforementioned architectures, each with varying training setups (see \Cref{tbl:vocoders}), and then investigate the fingerprints of various vocoders trained with different setups to analyze the influence of training setup on the fingerprint.

\subsection{Acoustic model fingerprint}\label{sec:method:acoustic}

Acoustic models are paired with a vocoder to generate waveforms, therefore, we study two aspects of the acoustic model fingerprint: when one vocoder is used to generate waveforms from the mel spectrograms produced by the acoustic model, and when multiple vocoders are used to generate waveforms from the mel spectrograms produced by the acoustic model. We train acoustic models of 3 architectures: Tacotron~2~\cite{shen2018tacotron2}, FastSpeech~2~\cite{ren2021fastspeech}, and GradTTS~\cite{popov2021gradtts}, and use our trained models in conjunction with the previously trained vocoders to generate waveforms from the mel spectrograms produced by the acoustic models. We then investigate the fingerprints of the generated waveforms to determine whether the acoustic model leaves a fingerprint.

\subsection{Relationship between fingerprints}\label{sec:fingerprint:model}

When it comes to the relationship between vocoder and acoustic model fingerprints, there are two competing hypotheses, both needing to be verified: either (1) vocoder fingerprints can visibly interfere with acoustic model fingerprints, or (2) acoustic model fingerprints can visibly interfere with vocoder fingerprints. To verify both hypotheses, we design two experiments with complementary setups, aimed to verify the two aforementioned, somewhat contradictory hypotheses. By setting one module --- acoustic model or vocoder, as the case may be --- to be known in the testing set, and testing the fingerprint against unknown vocoders or acooustic models, respectively, we can verify the two hypotheses. The setups of the two experiments are shown in \Cref{tbl:exp-setup}, in rows R1 and R2.

\subsection{Perturbations}

We furthermore seek to investigate the robustness of these fingerprints against perturbations in speech samples, namely noise, reverberation, and speed adjustment. To this end, we perform three experiments, for each of the aforementioned perturbations, where we add the specific perturbation to the speech samples in the test set, and then investigate the fingerprints of the generated waveforms to determine the extent of impact such perturbation would have on the original fingerprints. 
The setups of the three experiments are shown in \Cref{tbl:exp-setup}, in rows N1, N2 and N3.

\section{Experiments and results}

In this section, we outline our conducted experiments and the subsequent findings. We have undertaken a series of ten distinct experiments. The first five experiments are conducted to verify the existence of vocoder (V1--V3) and acoustic model (A1--A2) fingerprints in synthesized speech, and the next two (R1--R2) are conducted to verify the influence of the acoustic model and the vocoder on the fingerprints. Finally, three experiments (N1--N3) are conducted to investigate the robustness of the fingerprints against perturbations in the input speech samples. The setups of the experiments are shown in \Cref{tbl:exp-setup}.

\subsection{Experimental setup}

In the process of analyzing speech waveforms, each sample undergoes a detailed extraction method to obtain its linear frequency cepstral coefficients (LFCCs). This process begins with the segmentation of the speech waveform into frames, from which we extract 20 LFCCs for each frame. This extraction uses a Hamming window, which is a specific type of function used to smooth out the signal, minimizing the edge effects during the analysis. The parameters for this window are carefully chosen, with a window length set at \(M = 480\) and hop length of \(H = 240\). 

We standardize the size of the resulting LFCC matrix across all samples, since the matrix might vary in the number of frames depending on the duration of the speech sample. To address this variance, we either truncate the matrix to contain exactly 500 frames if it exceeds this number, or we pad it with zeros to reach the frame count of 500 if it contains fewer frames. This step is crucial for maintaining consistency across the dataset, allowing the neural network to process the data more effectively.

Inspired by \newcite{marra2019gans}, the LFCC representation of the speech sample is then passed to a Res2Net network~\cite{gao2021res2net} and projected onto a 256-dimension feature space, then passed through a single fully-connected (FC) layer classifier to decide between the finite number of sources. The entire network, i.e., both the extraction network and the classifier, is trained using cross-entropy loss:
\begin{equation}
  \mathcal{L} = -\sum_{i=1}^{N} \sum_{j=1}^{C} y_{i}^{(j)} \log(p_{i}^{(j)})
\end{equation}
where \(N\) is the number of samples, \(C\) is the number of classes, \(y\) and \(p\) denote the ground truth label and the predicted probability, respectively.

\subsection{Dataset}

In our experiments, we use the multi-speaker dataset LibriTTS~\cite{zen2019libritts} to both train acoustic models and vocoders as well as to generate the waveform corpus for fingerprint extraction. The ``clean'' training set of LibriTTS, containing 160,369 genuine utterances from 1,230 speakers, totaling 262 hours of speech, is used to train all models until convergence.

In departure from the traditional setup where the validation set and training set are from the same distribution, we synthesize the validation set using the same setups as the test set (see \Cref{tbl:exp-setup}), since we are interested in the performance of the classifier on the test set, which is synthesized using different setups from the training set. The training, validation and test sets are, however, disjoint, to help avoid overfitting.

\begin{table}
	\centering
	\small
	\caption{Summary of experiment results (\%)}\label{tbl:exp:summary}
	\begin{tabularx}{0.55\linewidth}{lcRRR}
		\toprule
		\textbf{Variable} & \textbf{Exp.} & \textbf{Precision} & \textbf{Recall} & \textbf{F}$_{\textbf{1}}$ \\
		\midrule
		\multirow{3}{*}{Vocoder} & V1 & 98.46 & 98.44 & 98.44 \\
		& V2 & 95.74 & 95.73 & 95.73 \\
		& V3 & 91.57 & 91.19 & 91.20 \\ \midrule
		\multirow{2}{*}{Acoustic model} & A1 & 99.79 &  99.79 & 99.79\\ 
		& A2 & 98.99 & 98.99 & 98.99\\
		\midrule
		\multirow{2}{*}{Relationship} & R1 & 23.75 & 25.01 & 10.05 \\
		& R2 & 98.46 &98.42 & 98.42 \\ 
		\midrule
		\multirow{3}{*}{Perturbation} & N1 & 98.21 & 98.16 & 98.17 \\
		& N2 & 87.82 & 86.36 & 86.03 \\
		& N3 & 91.24 & 91.12 & 91.07 \\
		\bottomrule
	\end{tabularx}
\end{table}

\begin{figure*}[t]
	\centering
	\makebox[\textwidth]{\makebox[1.0\textwidth]{
		\hspace*{\fill}
	\begin{subfigure}[T]{0.25\linewidth}
		\centering
		\includegraphics[width=\linewidth]{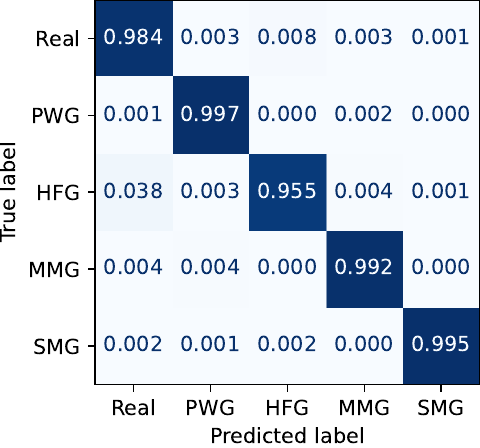}
		\caption{V1}
		\label{fig:vocexp:res2net-confusion-1}
	\end{subfigure}
	\hfill
	\begin{subfigure}[T]{0.23\linewidth}
		\centering
		\includegraphics[width=\linewidth]{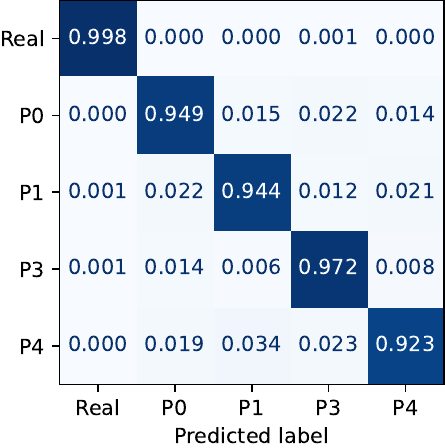}
		\caption{V2}
		\label{fig:vocexp:res2net-confusion-2}
	\end{subfigure}
	\hfill
	\begin{subfigure}[T]{0.23\linewidth}
		\centering
		\includegraphics[width=\linewidth]{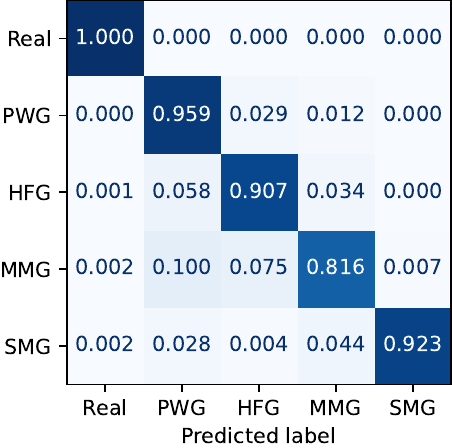}
		\caption{V3}
		\label{fig:vocexp:res2net-confusion-3}
	\end{subfigure}
	\hspace*{\fill}}}

	\vspace*{10pt}
	
	\makebox[\textwidth]{\makebox[1.0\textwidth]{
	\hspace*{\fill}
	\begin{subfigure}[T]{0.23\linewidth}
		\centering
		\includegraphics[width=\linewidth]{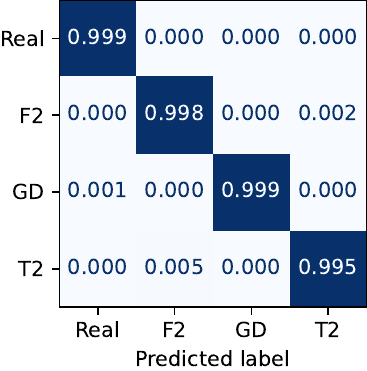}
		\caption{A1}
		\label{fig:ttsexp:res2net-confusion-1}
	\end{subfigure}
	\hspace*{\fill}
	\begin{subfigure}[T]{0.23\linewidth}
		\centering
		\includegraphics[width=\linewidth]{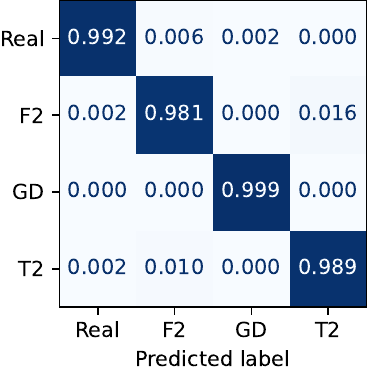}
		\caption{A2}
		\label{fig:ttsexp:res2net-confusion-2}
	\end{subfigure}
	\hspace*{\fill}
	\begin{subfigure}[T]{0.23\linewidth}
		\centering
		\includegraphics[width=\linewidth]{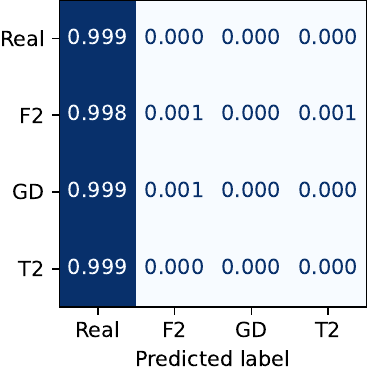}
		\caption{R1}
		\label{fig:relexp:res2net-confusion-1}
	\end{subfigure}
	\hspace*{\fill}}}

	\vspace*{10pt}

	\makebox[\textwidth]{\makebox[1.0\textwidth]{
	\hspace*{\fill}
	\begin{subfigure}[T]{0.23\linewidth}
		\centering
		\includegraphics[width=\linewidth]{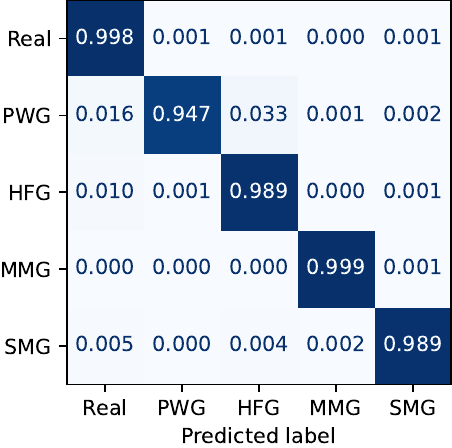}
		\caption{R2}
		\label{fig:relexp:res2net-confusion-2}
	\end{subfigure}
	\hspace*{\fill}
	\begin{subfigure}[T]{0.24\linewidth}
		\centering
		\includegraphics[width=\linewidth]{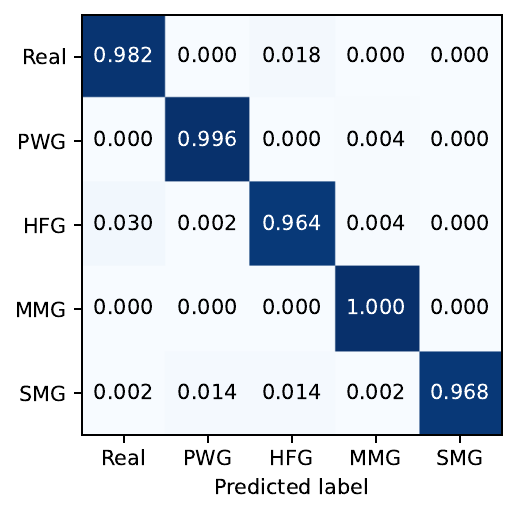}
		\caption{N1}
		\label{fig:pertexp:res2net-confusion-1}
	\end{subfigure}
	\hspace*{\fill}
	\begin{subfigure}[T]{0.24\linewidth}
		\centering
		\includegraphics[width=\linewidth]{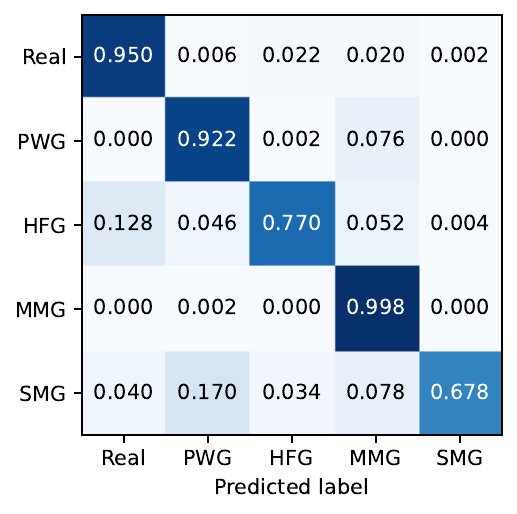}
		\caption{N2}
		\label{fig:pertexp:res2net-confusion-2}
	\end{subfigure}
	\hspace*{\fill}
	\begin{subfigure}[T]{0.24\linewidth}
		\centering
		\includegraphics[width=\linewidth]{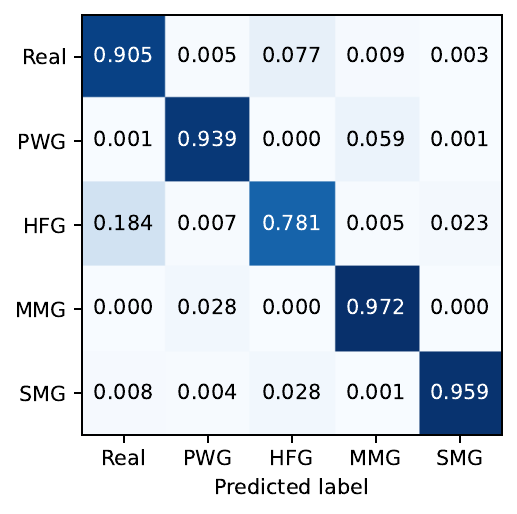}
		\caption{N3}
		\label{fig:pertexp:res2net-confusion-3}
	\end{subfigure}
	\hspace*{\fill}}}
	\caption{Confusion matrices of experiments in our study.}\label{fig:relation:res2net-confusion}
\end{figure*}

\subsection{Evaluation metric}

The performance of fingerprint extraction and analysis should align with the ability to classify the audio samples and identify their source of generation. To thoroughly evaluate the performance of our fingerprint extraction model, we employ a trio of established metrics commonly utilized in classification tasks: precision, recall, and the F1 score. These metrics offer a comprehensive view of the model's accuracy and its ability to distinguish between different sources of generated speech.
Beyond these numerical metrics, we delve into the analysis of the confusion matrix associated with our classifier, to gain insight into the performance of the fingerprint extraction model, as it visualizes the performance of a classifier by identifying the number of instances where the classifier misclassified the speech samples and in turn evaluate the performance of the fingerprint extraction model, on a per-category basis. This allows us to go beyond the global metrics and gain a more nuanced understanding of the model's performance.

\subsection{Results and analysis}

Following the methodology outlined in Section~2, and the setup of the experiments in \Cref{tbl:exp-setup}, we conduct a series of experiments, aimed at verifying the existence of vocoder and acoustic model fingerprints in synthesized speech, and to investigate the relationship between the two fingerprints. We also investigate the robustness of the fingerprints against perturbations in the input speech samples. The results of these experiments are summarized in \Cref{tbl:exp:summary}, and the confusion matrices of the experiments are shown in \Cref{fig:relation:res2net-confusion}.

\subsubsection{Vocoder fingerprints}\label{sec:vocoder}

For vocoder fingerprint analysis, we use copy synthesis to emulate a fixed acoustic model, minimizing variables. We train 3 vocoders for each of the four architectures: Parallel WaveGAN (PWG), HiFiGAN (HFG), Multiband MelGAN (MMG), and Style MelGAN (SMG), each with varying seeds and batch sizes (see \Cref{tbl:vocoders}). We additionally train two more PWG vocoders with different initialization seeds to investigate the impact of training setup on fingerprints. Three experiments assess vocoder architecture and training setup impact on fingerprints. In V1 and V3, we apply copy synthesis on the training set using P0, H0, and M0 models, labeling results by vocoder architecture for Res2Net classifier training. The validation and testing sets differ in vocoder models used for copy synthesis (see \Cref{tbl:exp-setup}). In V2 we investigate training setup impact by classifying along initialization seed lines within the same architecture (i.e., Parallel WaveGAN).

Results are summarized in the V1--V3 rows of \Cref{tbl:exp:summary} as well as the confusion matrices in \Crefrange{fig:vocexp:res2net-confusion-1}{fig:vocexp:res2net-confusion-3}. The high F1 score of V1 and the highly diagonal confusion matrix in \Cref{fig:vocexp:res2net-confusion-1} confirm architecture-specific fingerprints in the generated waveforms serving as strong indicators of source. In contrast, the slight degredation of classification performance in Experiment V3, as shown in \Cref{fig:vocexp:res2net-confusion-3}, suggests vocoder training setups weaken fingerprints, an observation corroborated by Experiment V2 (\Cref{fig:vocexp:res2net-confusion-2}). Overall, results confirm the existence of finer-grained vocoder fingerprints based on parameter differences, but with less pronounced impact than architectural ones.

\subsubsection{Acoustic model fingerprints}\label{sec:exp:acoustic}

We investigate and analyze the fingerprint of the acoustic model. Three models, Tacotron~2, FastSpeech~2 and Grad-TTS, are trained on the LibriTTS dataset. These models generate labeled speech waveforms from the LibriTTS training set text corpus. We use the entire speech corpus, encompassing both real and synthesized samples, to train the Res2Net classifier. The same approach applies to validation and test sets. To thoroughly examine the detectability and characteristics of the acoustic model fingerprint, we conduct two distinct experiments. Experiment A1 serves as a baseline, where we limit the diversity of the vocoders used in the synthesis process. Experiment A2, on the other hand, introduces a wider variety of vocoders into the mix (see \Cref{tbl:exp-setup}). This variation allows us to assess the impact of vocoder diversity on the ability of the Res2Net classifier to learn and identify acoustic model fingerprints, and more importantly, to ascertain that the fingerprint features learned by our model are indeed acoustic model-specific.

The results of our study are encapsulated in confusion matrices, as referenced in our report (\Crefrange{fig:ttsexp:res2net-confusion-1}{fig:ttsexp:res2net-confusion-2}), and the detailed analysis is provided in rows A1--A2 of \Cref{tbl:exp:summary}. These results shed light on our question, demonstrating that the acoustic model fingerprint is indeed detectable with a high degree of accuracy when the vocoder used during the synthesis process is either fixed or known. This revelation underscores the effectiveness of a fixed vocoder setup in enhancing the learning of acoustic model fingerprints. However, it is also noteworthy that introducing a variance in the vocoders, as done in Experiment A2, does not significantly impede the learning process. This is particularly true when the vocoder variance is evenly distributed across the categories under investigation. The insights gained from Experiment A2 highlight the resilience of the acoustic model fingerprint learning process, even in the face of increased diversity in vocoder usage.

\subsubsection{Relationship between fingerprints}\label{sec:exp:relationship}

As outlined in Section~\ref{sec:fingerprint:model}, we investigate the relationship between vocoder and acoustic model fingerprints with a pair of complementary experiments, R1 and R2. In R1, we fix the acoustic model and vary the vocoder; the acoustic models used to generate the testing set are known, but the vocoder (H0) is swapped for two unknown models (P1 and S1). In R2, we fix the vocoder architectures, but vary the acoustic models used to generate the testing set (TN being replaced by FS). The results of these experiments are summarized in the R1--R2 rows of \Cref{tbl:exp:summary} and the confusion matrices in \Crefrange{fig:relexp:res2net-confusion-1}{fig:relexp:res2net-confusion-2}. Results in \Cref{fig:relexp:res2net-confusion-1} demonstrate that when vocoders vary, the acoustic model fingerprint becomes nearly undetectable, aligning with our hypothesis that vocoder fingerprints overshadow acoustic model fingerprints. This is further supported by the results in \Cref{fig:relexp:res2net-confusion-2}, where the vocoder fingerprint is still detectable when the acoustic model varies.

The experiment outcomes align with our expectations, considering the distinct roles of the vocoder and acoustic model in the process of speech synthesis. The vocoder directly generates the waveform by translating the spectrogram produced by the acoustic model into speech. Given its role as the last step in the synthesis chain, the vocoder's involvement in the generation of the speech waveform is direct and substantial. It takes the more compact, frequency-based information from the spectrogram and transforms it into the waveform that we ultimately hear as speech. This direct engagement implies that it has a significant influence on the final characteristics of the speech waveform. Consequently, it's plausible to suggest that the vocoder leaves a more distinguishable fingerprint on the generated speech than the acoustic model does.

This distinction between the vocoder and acoustic model's contributions is essential for understanding how different components of the speech synthesis process affect the final product. It implies that when analyzing synthesized speech for the purpose of identifying its source or assessing its quality, the vocoder's characteristics may be a more telling factor than those of the acoustic model. This insight is crucial for the development of speech synthesis systems and the evaluation of their performance.

\subsection{Fingerprint robustness against speech perturbation}

Given the analyses above, demonstrating the existence of distinct fingerprints for both vocoders and acoustic models, as well as the overshadowing of acoustic model fingerprints by vocoder fingerprints, we further investigate the robustness of the fingerprints against perturbations in the input speech samples. We analyze the robustness from three aspects:
\begin{enumerate}[nosep]
	\item Noise: We add white noise to the speech samples in the test set of R2, with various signal-to-noise ratios (SNR) of 0--10 dB.
	\item Reverberation: We add reverberation to the speech samples in the test set of R2, with various reverberation times (RT) of 0.5--1.5 seconds.
	\item Speed adjustment: We adjust the speed of the speech samples in the test set of R2, with various speed factors of 0.9x--1.1x.
\end{enumerate}

We conduct an experiment where we add noise to the speech samples in the test set. We use SoX\footnote{Available at \url{https://sourceforge.net/projects/sox/}} to add white noise to the speech samples in the test set of R2, with a signal-to-noise ratio (SNR) of 10 dB. We then investigate the fingerprints of the generated waveforms to determine the extent of impact such perturbation would have on the original fingerprints, as summarized in Row N1 of \Cref{tbl:exp:summary}.

\begin{table}
	\centering
	\small
	\caption{Summary of perturbation experiment results (\%)}\label{tbl:exp-pert:summary}
	\begin{tabularx}{0.65\linewidth}{lXRRR}
		\toprule
		\textbf{Variable} & \textbf{Param.} & \textbf{Precision} & \textbf{Recall} & \textbf{F}$_{\textbf{1}}$ \\
		\midrule
		\multirow{2}{*}{Noise} & SNR=10dB & 98.21 & 98.16 & 98.17 \\
		& SNR=5dB & 97.04 & 96.98 & 97.00 \\
		\midrule
		\multirow{2}{*}{Reverberation} & RT=0.5s & 87.82 & 86.36 & 86.03 \\
		& RT=1.0s & 85.12 & 83.68 & 83.35 \\
		\midrule
		\multirow{2}{*}{Speed adjustment} & 0.9x & 91.24 & 91.12 & 91.07 \\
		& 1.1x & 90.12 & 89.98 & 89.94 \\
		\bottomrule
	\end{tabularx}
\end{table}

Similar setups are used to investigate the robustness of the fingerprint extraction model against reverberations and speed adjustments (varying from 0.9x to 1.1x), with results listed in rows N2 and N3 of \Cref{tbl:exp:summary}, respectively. The results of these experiments show that the fingerprint extraction model is somewhat robust against perturbations in the input speech samples, with a slight degradation in performance. Among the three perturbations tested, reverberations seem to have the most adverse effect on the fingerprints. This is likely due to the fact that reverberations are more likely to affect the spectral characteristics of the speech samples, which are the basis of the fingerprints, than noise and speed adjustments.

We furthermore conduct experiments with various signal-to-noise ratios (SNR) and reverberation times (RT) to investigate the robustness of the fingerprints against noise and reverberation. The results of these experiments are summarized in \Cref{tbl:exp-pert:summary}. The results reinforce our analysis of the fingerprint robustness against perturbations, with speed adjustments and reverberation having a more pronounced impact on the fingerprints than noise, with reverberation being the most detrimental to the fingerprints.

\begin{figure}
\begin{subfigure}{.4\linewidth}
	\centering
	\includegraphics[width=\linewidth]{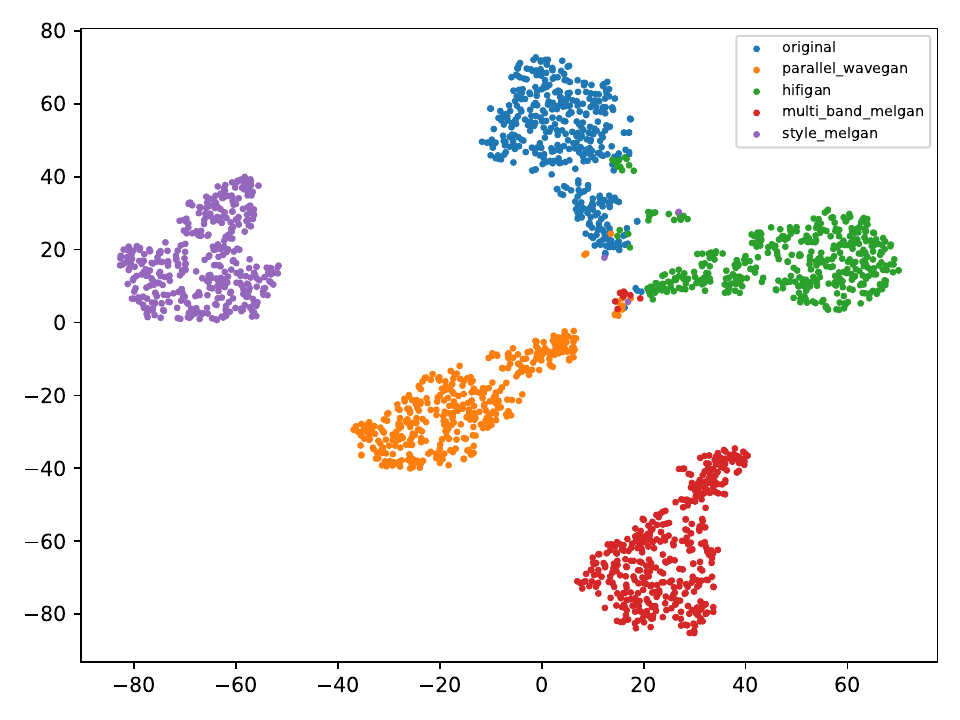}
	\caption{Vocoder fingerprints (V3)}
	\label{fig:tsne:vocoder}
\end{subfigure}
\hfill
\begin{subfigure}{.4\linewidth}
	\centering
	\includegraphics[width=\linewidth]{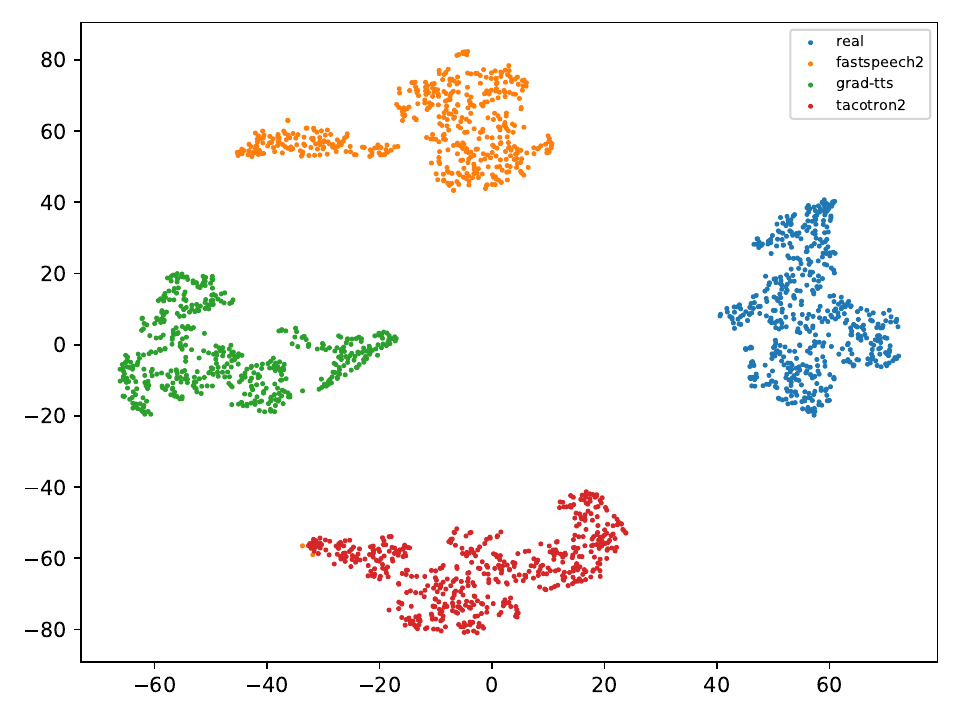}
	\caption{Acoustic model fingerprints (A2)}
	\label{fig:tsne:acoustic}
\end{subfigure}
\hfill
\begin{subfigure}{.4\linewidth}
	\centering
	\includegraphics[width=\linewidth]{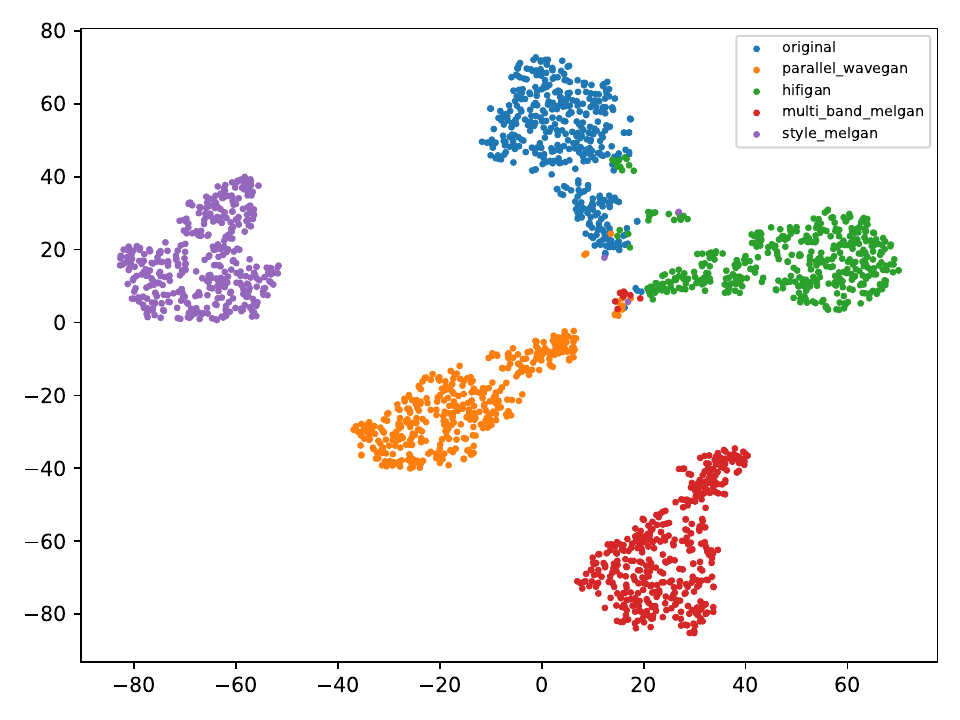}
	\caption{Relationship between vocoder and acoustic model fingerprints (R2)}
	\label{fig:tsne:relationship}
\end{subfigure}
\hfill
\begin{subfigure}{.4\linewidth}
	\centering
	\includegraphics[width=\linewidth]{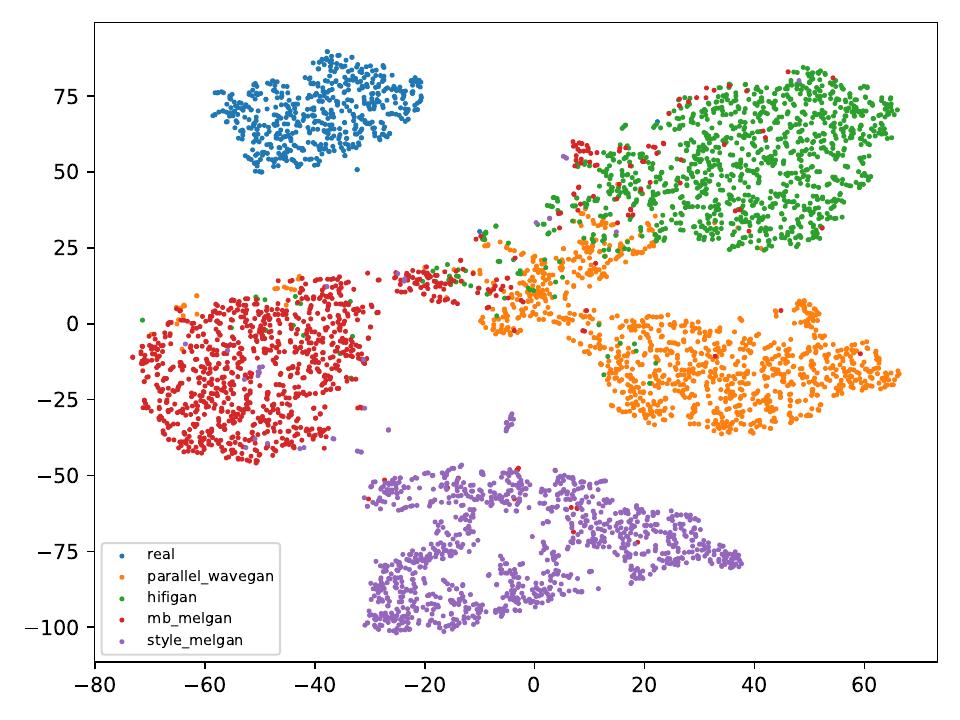}
	\caption{Perturbation experiment: audio speed readjustment (N3)}
	\label{fig:tsne:pert}
\end{subfigure}
\caption{t-SNE representation of fingerprints extracted in our experiments.}
\label{fig:tsne}
\end{figure}

\subsection{Visualization}

Given the results reported in the previous section, to better and more intuitively understand the fingerprints of the vocoders and acoustic models, we visualize the fingerprints of the vocoders and acoustic models using t-SNE~\cite{maaten2008visualizing}. 

We focus first on the vocoders. By applying t-SNE, we project the high-dimensional fingerprints of various vocoders onto a two-dimensional plane. This visualization is reflected in \Cref{fig:tsne:vocoder}. The resulting projection clearly illustrates how the fingerprints of the vocoders are dispersed, forming clearly distinguishable clusters on the plane. Each of these clusters represents a unique vocoder architecture, unequivocally demonstrating the presence of distinct, identifiable fingerprints for each vocoder.

Following the analysis of vocoders, we extend the same t-SNE visualization technique to the acoustic models. The outcomes of this process are captured in \Cref{fig:tsne:acoustic}, where, akin to the vocoders, the fingerprints of the acoustic models are also neatly organized into discernible clusters. This pattern reinforces the premise that acoustic models, much like vocoders, possess their own unique fingerprints. The clarity with which these fingerprints can be distinguished affirms the potential for precise identification and differentiation of acoustic models based on their inherent characteristics.

\Cref{fig:tsne:relationship} shows the t-SNE projection of the fingerprints of the vocoders and acoustic models in Experiment R2. As with the previous visualizations, the fingerprints of the vocoders are clearly separated into distinct clusters, despite the presence of different and unknown acoustic models. This result confirms that the vocoder fingerprints are more prominent than the acoustic model fingerprints, potentially overshadowing the acoustic model fingerprints.

Lastly, \Cref{fig:tsne:pert} shows the t-SNE projection of the fingerprints of the perturbation experiment N3. The fingerprints of the perturbed speech samples are still distinguishable, albeit with a slight overlap, indicating that the fingerprint extraction model is somewhat robust against perturbations in the input speech samples.

\section{Conclusion}

In conclusion, our study centers on extracting fingerprints from acoustic models and vocoders to identify the sources of synthesized speech. Our experiments confirm the existence of distinct, model-specific fingerprints for both vocoders and acoustic models. Notably, we observe that vocoders tend to leave more prominent fingerprints, potentially overshadowing those of the acoustic models. This discovery holds promise, indicating the feasibility of employing fingerprint extraction for source attribution in synthesized speech. This insight is particularly relevant for researchers and developers in the field of computational linguistics and speech synthesis, as it highlights the importance of carefully selecting or designing vocoders to achieve the desired qualities in synthesized speech. Nonetheless, our current research leaves room for further exploration. We have yet to consider the potential impact of different languages, which entail diverse phonological feature distributions, on the fingerprints of generated waveforms. It will also be interesting and important to devise methods for extracting fingerprints from end-to-end TTS systems, which are gaining traction in recent years. Lastly, even in the context of pipeline TTS systems, we have yet to come up with a reliable method for extracting acoustic-model-specific fingerprints despite the masking presence of vocoder fingerprints. These unexplored dimensions will form the focal point of our future investigations, seeking to provide a more comprehensive understanding of the nuances in the fingerprints of synthesized speech.

\section*{Acknowledgements}

This work is supported by the National Natural Science Foundation of China (NSFC) (No.~62322120, No.~62306316, No~U21B2010, No.~62206278), as well as the ANSO Scholarship for Young Talents.

\bibliographystyle{ccl}
\bibliography{refs}

\begin{thebibliography}{}

\bibitem[\protect\citename{Brock \bgroup et al.\egroup }2019]{brock2019large}
Andrew Brock, Jeff Donahue, and Karen Simonyan.
\newblock 2019.
\newblock Large scale {GAN} training for high fidelity natural image synthesis.
\newblock In {\em International Conference on Learning Representations}.

\bibitem[\protect\citename{Gao \bgroup et al.\egroup }2021]{gao2021res2net}
Shang-Hua Gao, Ming-Ming Cheng, Kai Zhao, Xin-Yu Zhang, Ming-Hsuan Yang, and Philip Torr.
\newblock 2021.
\newblock {Res2Net}: A new multi-scale backbone architecture.
\newblock {\em IEEE Transactions on Pattern Analysis and Machine Intelligence}, 43(2):652--662.

\bibitem[\protect\citename{Kim \bgroup et al.\egroup }2021]{kim2021vits}
Jaehyeon Kim, Jungil Kong, and Juhee Son.
\newblock 2021.
\newblock Conditional variational autoencoder with adversarial learning for end-to-end text-to-speech.
\newblock In {\em Proceedings of the 38th International Conference on Machine Learning}, page 5530–5540. PMLR.

\bibitem[\protect\citename{Kinnunen \bgroup et al.\egroup }2017]{kinnunen2017asvspoof}
Tomi Kinnunen, Md. Sahidullah, Héctor Delgado, Massimiliano Todisco, Nicholas Evans, Junichi Yamagishi, and Kong~Aik Lee.
\newblock 2017.
\newblock The {ASVspoof} 2017 challenge: Assessing the limits of replay spoofing attack detection.
\newblock In {\em Interspeech 2017}, page 2–6. ISCA.

\bibitem[\protect\citename{Kong \bgroup et al.\egroup }2020]{kong2020hifigan}
Jungil Kong, Jaehyeon Kim, and Jaekyoung Bae.
\newblock 2020.
\newblock Hifi-gan: Generative adversarial networks for efficient and high fidelity speech synthesis.
\newblock In {\em Advances in Neural Information Processing Systems}, volume~33, pages 17022--17033.

\bibitem[\protect\citename{Lu \bgroup et al.\egroup }2023]{lu2023detecting}
Jingze Lu, Yuxiang Zhang, Zhuo Li, Zengqiang Shang, WenChao Wang, and Pengyuan Zhang.
\newblock 2023.
\newblock Detecting unknown speech spoofing algorithms with nearest neighbors.
\newblock In {\em Proceedings of IJCAI 2023 Workshop on Deepfake Audio Detection and Analysis}.

\bibitem[\protect\citename{Marra \bgroup et al.\egroup }2019]{marra2019gans}
Francesco Marra, Diego Gragnaniello, Luisa Verdoliva, and Giovanni Poggi.
\newblock 2019.
\newblock Do {{GANs Leave Artificial Fingerprints}}?
\newblock In {\em 2019 {{IEEE Conference}} on {{Multimedia Information Processing}} and {{Retrieval}} ({{MIPR}})}, pages 506--511. {IEEE Computer Society}.

\bibitem[\protect\citename{Mustafa \bgroup et al.\egroup }2021]{mustafa2021stylemelgan}
Ahmed Mustafa, Nicola Pia, and Guillaume Fuchs.
\newblock 2021.
\newblock Stylemelgan: An efficient high-fidelity adversarial vocoder with temporal adaptive normalization.
\newblock In {\em 2021 IEEE International Conference on Acoustics, Speech and Signal Processing (ICASSP)}, pages 6034--6038.

\bibitem[\protect\citename{Pons \bgroup et al.\egroup }2021]{pons2021upsampling}
Jordi Pons, Santiago Pascual, Giulio Cengarle, and Joan Serrà.
\newblock 2021.
\newblock Upsampling artifacts in neural audio synthesis.
\newblock In {\em 2021 IEEE International Conference on Acoustics, Speech and Signal Processing (ICASSP)}, pages 3005--3009.

\bibitem[\protect\citename{Popov \bgroup et al.\egroup }2021]{popov2021gradtts}
Vadim Popov, Ivan Vovk, Vladimir Gogoryan, Tasnima Sadekova, and Mikhail Kudinov.
\newblock 2021.
\newblock {Grad-TTS}: A diffusion probabilistic model for text-to-speech.
\newblock In {\em Proceedings of the 38th International Conference on Machine Learning}, page 8599–8608. PMLR.

\bibitem[\protect\citename{Ren \bgroup et al.\egroup }2021]{ren2021fastspeech}
Yi~Ren, Chenxu Hu, Xu~Tan, Tao Qin, Sheng Zhao, Zhou Zhao, and Tie-Yan Liu.
\newblock 2021.
\newblock {FastSpeech} 2: Fast and high-quality end-to-end text to speech.
\newblock In {\em International Conference on Learning Representations (ICLR)}.

\bibitem[\protect\citename{Shen \bgroup et al.\egroup }2018]{shen2018tacotron2}
Jonathan Shen, Ruoming Pang, Ron~J. Weiss, Mike Schuster, Navdeep Jaitly, Zongheng Yang, Zhifeng Chen, Yu~Zhang, Yuxuan Wang, Rj~{Skerrv-Ryan}, Rif~A. Saurous, Yannis Agiomvrgiannakis, and Yonghui Wu.
\newblock 2018.
\newblock Natural {TTS} synthesis by conditioning {Wavenet} on {Mel} spectrogram predictions.
\newblock In {\em 2018 IEEE International Conference on Acoustics, Speech and Signal Processing (ICASSP)}, pages 4779--4783.

\bibitem[\protect\citename{Tian \bgroup et al.\egroup }2023]{tian2023deepfake}
Ye~Tian, Yunkun Chen, Yuezhong Tang, and Boyang Fu.
\newblock 2023.
\newblock Deepfake algorithm recognition through multi-model fusion based on manifold measure.
\newblock In {\em Proceedings of IJCAI 2023 Workshop on Deepfake Audio Detection and Analysis}.

\bibitem[\protect\citename{Todisco \bgroup et al.\egroup }2018]{todisco2018integrated}
Massimiliano Todisco, Héctor Delgado, Kong~Aik Lee, Md~Sahidullah, Nicholas Evans, Tomi Kinnunen, and Junichi Yamagishi.
\newblock 2018.
\newblock Integrated presentation attack detection and automatic speaker verification: Common features and gaussian back-end fusion.
\newblock In {\em Proc. Interspeech 2018}, pages 77--81.

\bibitem[\protect\citename{{van der Maaten} and Hinton}2008]{maaten2008visualizing}
Laurens {van der Maaten} and Geoffrey Hinton.
\newblock 2008.
\newblock Visualizing {{Data}} using t-{{SNE}}.
\newblock {\em Journal of Machine Learning Research}, 9(86):2579--2605.

\bibitem[\protect\citename{Wang \bgroup et al.\egroup }2020]{wang2020asvspoof}
Xin Wang, Junichi Yamagishi, Massimiliano Todisco, Héctor Delgado, Andreas Nautsch, Nicholas Evans, Md~Sahidullah, Ville Vestman, Tomi Kinnunen, Kong~Aik Lee, Lauri Juvela, Paavo Alku, Yu-Huai Peng, Hsin-Te Hwang, Yu~Tsao, Hsin-Min Wang, Sébastien~Le Maguer, Markus Becker, Fergus Henderson, Rob Clark, Yu~Zhang, Quan Wang, Ye~Jia, Kai Onuma, Koji Mushika, Takashi Kaneda, Yuan Jiang, Li-Juan Liu, Yi-Chiao Wu, Wen-Chin Huang, Tomoki Toda, Kou Tanaka, Hirokazu Kameoka, Ingmar Steiner, Driss Matrouf, Jean-François Bonastre, Avashna Govender, Srikanth Ronanki, Jing-Xuan Zhang, and Zhen-Hua Ling.
\newblock 2020.
\newblock {ASVspoof} 2019: a large-scale public database of synthesized, converted and replayed speech.
\newblock {\em Computer Speech \& Language}, 64:101--114.

\bibitem[\protect\citename{Wu \bgroup et al.\egroup }2015]{wu2015asvspoof}
Zhizheng Wu, Tomi Kinnunen, Nicholas Evans, Junichi Yamagishi, Cemal Hanilçi, Md. Sahidullah, and Aleksandr Sizov.
\newblock 2015.
\newblock {ASVspoof} 2015: the first automatic speaker verification spoofing and countermeasures challenge.
\newblock In {\em Interspeech 2015}, page 2037–2041. ISCA.

\bibitem[\protect\citename{Yamagishi \bgroup et al.\egroup }2021]{yamagishi2021asvspoof}
Junichi Yamagishi, Xin Wang, Massimiliano Todisco, Md~Sahidullah, Jose Patino, Andreas Nautsch, Xuechen Liu, Kong~Aik Lee, Tomi Kinnunen, Nicholas Evans, et~al.
\newblock 2021.
\newblock Asvspoof 2021: accelerating progress in spoofed and deepfake speech detection.
\newblock {\em arXiv preprint arXiv:2109.00537}.

\bibitem[\protect\citename{Yamamoto \bgroup et al.\egroup }2020]{yamamoto2020parallel}
Ryuichi Yamamoto, Eunwoo Song, and Jae-Min Kim.
\newblock 2020.
\newblock Parallel wavegan: A fast waveform generation model based on generative adversarial networks with multi-resolution spectrogram.
\newblock In {\em 2020 IEEE International Conference on Acoustics, Speech and Signal Processing (ICASSP)}, pages 6199--6203.

\bibitem[\protect\citename{Yan \bgroup et al.\egroup }2022]{yan2022initial}
Xinrui Yan, Jiangyan Yi, Jianhua Tao, Chenglong Wang, Haoxin Ma, Tao Wang, Shiming Wang, and Ruibo Fu.
\newblock 2022.
\newblock An initial investigation for detecting vocoder fingerprints of fake audio.
\newblock In {\em Proceedings of the 1st International Workshop on Deepfake Detection for Audio Multimedia (DDAM '22)}, page 61–68.

\bibitem[\protect\citename{Yang \bgroup et al.\egroup }2020]{yang2020multiband}
Geng Yang, Shan Yang, Kai Liu, Peng Fang, Wei Chen, and Lei Xie.
\newblock 2020.
\newblock Multi-band {{MelGAN}}: Faster waveform generation for high-quality text-to-speech.
\newblock {\em arXiv:2005.05106 [cs, eess]}.

\bibitem[\protect\citename{Yi \bgroup et al.\egroup }2022]{yi2022add}
Jiangyan Yi, Ruibo Fu, Jianhua Tao, Shuai Nie, Haoxin Ma, Chenglong Wang, Tao Wang, Zhengkun Tian, Ye~Bai, Cunhang Fan, et~al.
\newblock 2022.
\newblock {ADD} 2022: the first audio deep synthesis detection challenge.
\newblock In {\em 2022 IEEE International Conference on Acoustics, Speech and Signal Processing (ICASSP)}, pages 9216--9220. IEEE.

\bibitem[\protect\citename{Yi \bgroup et al.\egroup }2023a]{yi2023add}
Jiangyan Yi, Jianhua Tao, Ruibo Fu, Xinrui Yan, Chenglong Wang, Tao Wang, Chu~Yuan Zhang, Xiaohui Zhang, Yan Zhao, Yong Ren, Le~Xu, Junzuo Zhou, Hao Gu, Zhengqi Wen, Shan Liang, Zheng Lian, Shuai Nie, and Haizhou Li.
\newblock 2023a.
\newblock {ADD 2023}: the second audio deepfake detection challenge.
\newblock In {\em Proceedings of IJCAI 2023 Workshop on Deepfake Audio Detection and Analysis}.

\bibitem[\protect\citename{Yi \bgroup et al.\egroup }2023b]{yi2023audio}
Jiangyan Yi, Chenglong Wang, Jianhua Tao, Xiaohui Zhang, Chu~Yuan Zhang, and Yan Zhao.
\newblock 2023b.
\newblock Audio deepfake detection: A survey.
\newblock {\em arXiv 2308.14970}.

\bibitem[\protect\citename{Zen \bgroup et al.\egroup }2019]{zen2019libritts}
Heiga Zen, Viet Dang, Rob Clark, Yu~Zhang, Ron~J. Weiss, Ye~Jia, Zhifeng Chen, and Yonghui Wu.
\newblock 2019.
\newblock {LibriTTS}: A corpus derived from librispeech for text-to-speech.
\newblock In {\em Proc. Interspeech 2019}, pages 1526--1530.

\end{thebibliography}

\end{document}